# Influence of chemical interactions on the electronic properties of BiOI/organic semiconductor heterojunctions for application in solution-processed electronics

**Vaidehi Lapalikar**[a], **Preetam Dacha**[bc], **Mike Hambsch**[c], **Yvonne J. Hofstetter** [de], **Yana Vaynzof** [de], **Stefan C. B. Mannsfeld**\*[bc] and **Michael Ruck** \*[afg]

[a]Faculty of Chemistry and Food Chemistry, Technische Universität Dresden, 01062 Dresden, Germany. E-mail: michael.ruck@tu-dresden.de
[b]Faculty of Electrical and Computer Engineering, Technische Universität Dresden, 01069 Dresden, Germany
[c]Center for Advancing Electronics Dresden (cfaed), Technische Universität Dresden, 01062 Dresden, Germany
[d]Chair for Emerging Electronic Technologies, Technische Universität Dresden, Nöthnitzer Str. 61, 01187 Dresden, Germany
[e]Leibniz-Institute for Solid State and Materials Research Dresden, Helmholtzstraße 20, 01069 Dresden, Germany
[f]Max Planck Institute for Chemical Physics of Solids, Nöthnitzer Str. 40, 01187 Dresden, Germany
[g]Würzburg-Dresden Cluster of Excellence ct.qmat, Technische Universität Dresden, 01062 Dresden, Germany

Bismuth oxide iodide (BiOI) has been viewed as a suitable environmentally-friendly alternative to lead-halide perovskites for low-cost (opto-)electronic applications such as photodetectors, phototransistors and sensors. To enable its incorporation in these devices in a convenient, scalable, and economical way, BiOI thin films were investigated as part of heterojunctions with various p-type organic semiconductors (OSCs) and tested in a field-effect transistor (FET) configuration. The hybrid heterojunctions, which combine the respective functionalities of BiOI and the OSCs were processed from solution under ambient atmosphere. The characteristics of each of these hybrid systems were correlated with the physical and chemical properties of the respective materials using a concept based on heteropolar chemical interactions at the interface. Systems suitable for application in lateral transport devices were identified and it was demonstrated how materials in the hybrids interact to provide improved and synergistic properties. These indentified heterojunction FETs are a first instance of successful incorporation of solution-processed BiOI thin films in a three-terminal device. They show a significant threshold voltage shift and retained carrier mobility compared to pristine OSC devices and open up possibilities for future optoelectronic applications.

**Introduction**

Extensive research on lead-halide perovskites has established the possibility of high quality, up-scalable (opto-)electronics fabricated using solution processing and at the same time, highlighted their shortcomings owing to the toxicity and poor stability of perovskite-based devices.[1–3] Many of the compelling properties of lead-based perovskites can be traced back to the $ns^2$-configuration of the lead(II) cation.[4] Therefore, materials based on non-toxic bismuth(III), tin(II) or antimony(III) cations, which also have an $ns^2$ outer shell electronic configuration, have been viewed as the obvious next choices, with extensive work being invested in driving up the efficiencies of (opto-)electronic devices that are based on them.[5–

[7] Bismuth(III) presents a particularly appealing choice over lead(II) since it is an environmentally compatible, but heavy metal ion, thus allowing for spin–orbit coupling resulting in substantial defect tolerance due to mainly shallow defects.[8,9] The defect tolerance translates experimentally to the possibility of fast and solution-based processing. The most common classes of materials explored herein for electronic and optoelectronic applications are binary halides, ternary halides (perovskite-inspired materials), chalcogenides and chalcogenide halides.[10–13] Bismuth chalcogenide halides are an interesting class of semiconductors due to a combination of favorable features such as suitable bandgaps for light detection and harvesting, high light absorption coefficients, effective charge separation, and high chemical, thermal and, operational stability. For these reasons, they are being investigated extensively for applications in photocatalysis,[14,15] photovoltaics,[10,16–18] radiation detectors,[19] supercapacitors,[20–23] *etc*. Furthermore, they can be fabricated in a wide variety of synthetic routes, which facilitate tunability of morphology, optical properties, and ease of processing into functional systems.

An exemplary material from this class is that of bismuth oxide iodide (BiOI), which can be synthesized readily and has been largely studied for photocatalysis and photovoltaics in the form of bulk powder or thin films.[24–26] However, the anisotropic carrier mobility in BiOI combined with the highly textured morphology of its thin films present significant challenges to enhancing the performance of electronic devices based on BiOI. These challenges are frequently addressed by fabricating epitaxially grown thin films using vapor techniques in vacuum that result in oriented particles with better charge transport properties or by fabricating single-crystal devices.[27–31] These techniques are not only process intensive and require highly controlled fabrication conditions, but also rely on higher temperatures for vaporing the precursors and annealing of the resulting film to ensure optimal crystallinity. Given that BiOI is made of relatively inexpensive elements and can quite conveniently be obtained as high-quality thin films, it would be most beneficial to develop devices based on BiOI fabricated from solution in ambient air.

A particularly appealing strategy that could enable the integration of solution-processed BiOI in electronic devices could be to utilize it as part of a hybrid system, in specific where BiOI could bring in the functionality, while charge transport is carried out predominantly by a second material. This strategy has proven useful in employing many other emerging inorganic semiconductors in optoelectronic applications.[12,32–35] Thus, in this work we investigate – for the first time – heterojunctions of BiOI as two-component organic–inorganic hybrids that can be easily processed from solution in air and offer the possibility for scale-up. Such hybrid systems comprising of two different materials, have numerous benefits over single-component systems in terms of enhanced and added properties and customization options in addition to possible cost-effectiveness, and sustainability.[12,33–35] However, when fabricating functional hybrids, it is essential to discuss, in addition to the electronic interactions, the chemical interactions between the component layers in order to utilize them in electronic devices in the most suitable manner. Considering the extent to which chemical interactions can affect the electronic properties of the devices and the fact that these aspects have received little attention in most studies of hybrid devices, we focus our work on highlighting these aspects in detail.

To enable electronic devices of solution-processed BiOI thin films, we chose heterojunctions with organic semiconductors as these are soft materials and counteract the rigid and highly textured nature of the inorganic layer. We combine BiOI with p-type organic semiconductors (OSCs) to form hybrid systems as bilayer heterojunctions and study them as part of the bottom-gate bottom-contact (BGBC) field-effect transistor (FET) architecture. We have selected commonly used hydrophobic OSCs to enable layer-by-layer deposition, giving special considerations to design parameters such as solvents, annealing temperatures, *etc.* to ensure compatibility of processing of the two layers. The OSCs used here are the conjugated polymers **P3HT**, **PCDTPT**, **DPPDTT** and **PDPP4T** and the small-molecule OSC **TIPS-pentacene** (for formulas and full names see Fig. 1).

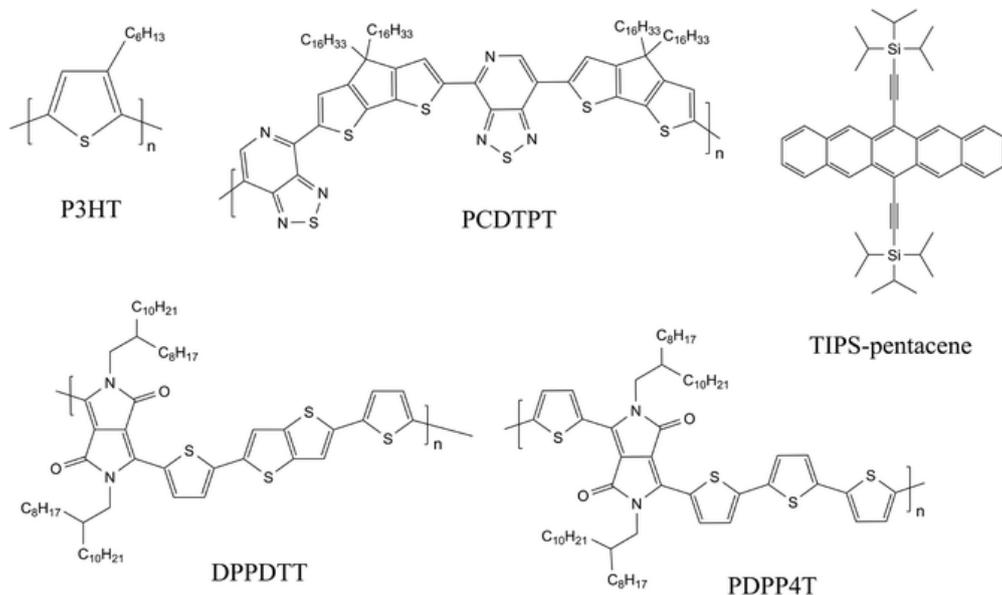

**Fig. 1** Molecular structures of OSCs **P3HT**: regioregular poly(3-hexylthiophene-2,5-diyl), **PCDTPT**: (4,4-dihex-adecyl-4*H*-cyclopenta[[1,2-*b*:5,4-b]dithiophen-2-yl)-*alt*-[1,2,5]thiadiazolo[3,4-*c*]pyridine], **TIPS-pentacene**: tri-isopropylsilyly-ethynyl, **DPPDTT**: poly[[2,3,5,6-tetrahydro-2,5-bis(2-octyldodecyl)-3,6-dioxopyrrolo[3,4-*c*]pyrrole-1,4-diyl]-2,5-thiophenediylthieno[3,2-*b*]thiophene-2,5-diyl-2,5-thiophenediyl], and **PDPP4T**: (poly[2,5-bis(2-octyldodecyl)pyrrolo[3,4-c]pyrrole-1,4(2*H*,5*H*)-dione-3,6-diyl)-*alt*-(2,2′;5′,2″;5″,2‴-quaterthiophen-5,5‴-diyl)].

We then characterize the pristine and heterojunction layers to, on the one hand, identify the key parameters of the OSC that have the most pronounced effect on the BiOI layer formation and on the other, explain the distinct device performance of each of the heterojunctions by correlating them with the chemical and physical properties of the constituting materials. These heterojunctions have the potential to be investigated in a wide range of electronic and optoelectronic devices, such as solar cells, photodetectors, (chemical) sensors, and phototransistors in the future.

**Results and discussion**

The most commonly reported methods for fabrication of BiOI thin films include the successive ionic layer adsorption and reaction (SILAR) method or vacuum techniques such as chemical vapor deposition (CVD) or aerosol-assisted CVD.[24,30,31,36] Although SILAR[24,36] involves processing from solution, it is not always suitable for formation of heterostructures for application in electronics as it involves extensive and repeated contact with solvents, often results in films with impurities, and requires additional chemicals such as mild acids that can interact chemically with other device layers. For these reasons and to utilize a simple, low-temperature and up-scalable process for the fabrication of BiOI thin films, we adopted the strategy of wet-chemical conversion of spin-coated $BiI_3$ films for the synthesis of BiOI. This was based on a previously reported process[37] with certain alterations in the processing parameters made to suit the subsequent direct application in electronic devices. Specifically, BiOI was obtained by conversion of spin-coated $BiI_3$ film to BiOI by hydrolysis in a methanol–water bath. This procedure is very fast (few minutes) and requires only moderate annealing temperatures (<150 °C). In comparison to the previously reported procedure, as mentioned in the experimental section, the duration for which the sample was held in the solvent bath was reduced and the annealing temperature was increased to ensure that most water, that was either in the film or adsorbed, was eliminated. This resulted in a visually uniform, orange-colored film.

**Characterization of BiOI thin film**

Powder X-ray diffraction (PXRD) analysis of the BiOI thin film was performed to check phase purity and gain information about film texture. The diffractogram confirms the formation of pure tetragonal BiOI (ICSD: 391354) without presence of the precursor $BiI_3$ or other crystalline by-products (Fig. 2a). BiOI displays a 2D layered structure consisting of $[Bi_2O_2]^{2+}$ layers alternating with two $I^-$ layers along the $c$-direction (Fig. 2b). The compound can also be regarded as a van-der-Waals compound formed by [$Bi_2O_2I_2$] layers. Like other compounds with layered structures that have stronger chemical bonds in the *ab*-plane than in the stacking direction, particle growth of BiOI proceeds preferentially in this plane and leads to platelet morphology with {001} as main faces.[31] The platelet habitus of the crystals generally causes a preferred orientation in powders when measured using a flat stage sample holder, and consequently an overemphasis of the intensities of the reflections of the 00*l* series in PXRD measured in the Bragg–Brentano geometry. Compared to that, we observe a substantially decreased relative intensity of the 00*l*-reflections, which was the first indication of the predominantly vertical growth of platelets in the film. The sharp reflections also substantiate the high crystallinity of BiOI so formed. Although PXRD categorically confirms the crystallinity of BiOI it provides only a partial understanding of the texture of the film. To further understand the interplay of preferred orientation in particle growth in combination with their alignment in the film because of the processing technique, we investigated the BiOI thin film using grazing incidence wide-angle X-ray scattering (GIWAXS) technique (Fig. 2c). While the 1D plot along $Q_{xy}$ resembles the PXRD pattern of the thin film as expected, the intensity of the 001 reflection has a clear maximum in the $Q_z$ direction, which confirms a preferred vertical growth of BiOI particles on the substrate (Fig. 2d). [38,39]

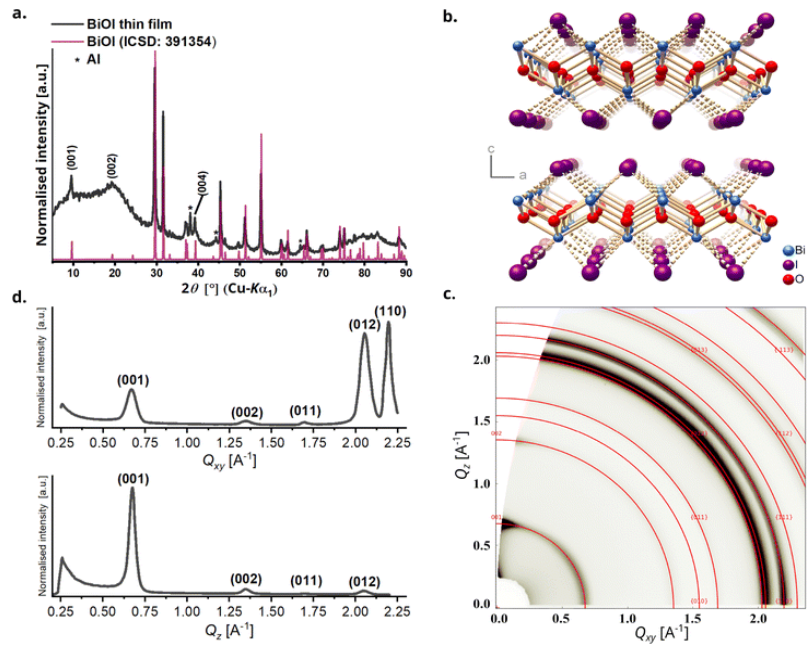

**Fig. 2** For BiOI, (a) PXRD of a thin film (diffuse scattering in the range $5° \leq 2\theta \leq 25°$ originates from the glass substrates used to grow the films and the reflections of aluminum originate from the sample holder), (b) crystal structure emphasizing the layers parallel to the *ab*-plane, (c) GIWAX reciprocal space map for thin film showing Debye–Scherrer rings marked with corresponding Miller indices, (d) 1D intensity profiles along $Q_{xy}$ and $Q_z$.

The morphology of the BiOI thin film was then studied using scanning electron microscopy (SEM) and atomic force microscopy (AFM) and the bulk composition using energy dispersive X-ray (EDX) analysis. The top-view SEM image, as also the AFM image, revealed a densely packed BiOI film with a highly textured morphology (Fig. 3a and b). As deduced from the X-ray analyses, the cross-sectional SEM image revealed a predominantly vertical alignment of BiOI particles on the substrate with a typical layer thickness of about 300 nm (Fig. 3c), which is in accordance with the thickness measured using profilometry. The SEM image of the precursor $BiI_3$ thin film (Fig. S1, ESI†) shows a compact layer of $BiI_3$. It is evident from this that the morphology of the $BiI_3$ layer does not template the resulting BiOI thin film. The bulk composition of the BiOI thin film was measured using EDX spectroscopy with measurements made at multiple locations on the film. It revealed an average atomic ratio Bi:I of 1:0.97(2). With respect to the accuracy of the method and systematical errors (*e.g.*, sample preparation), an iodine deficiency cannot be deduced from this result. The values are in accordance with previously reported values on solution processed thin films of BiOI.[40] A representative EDX spectrum of the thin film on Si wafer shows peaks of Bi, I and O (Fig. 3d).

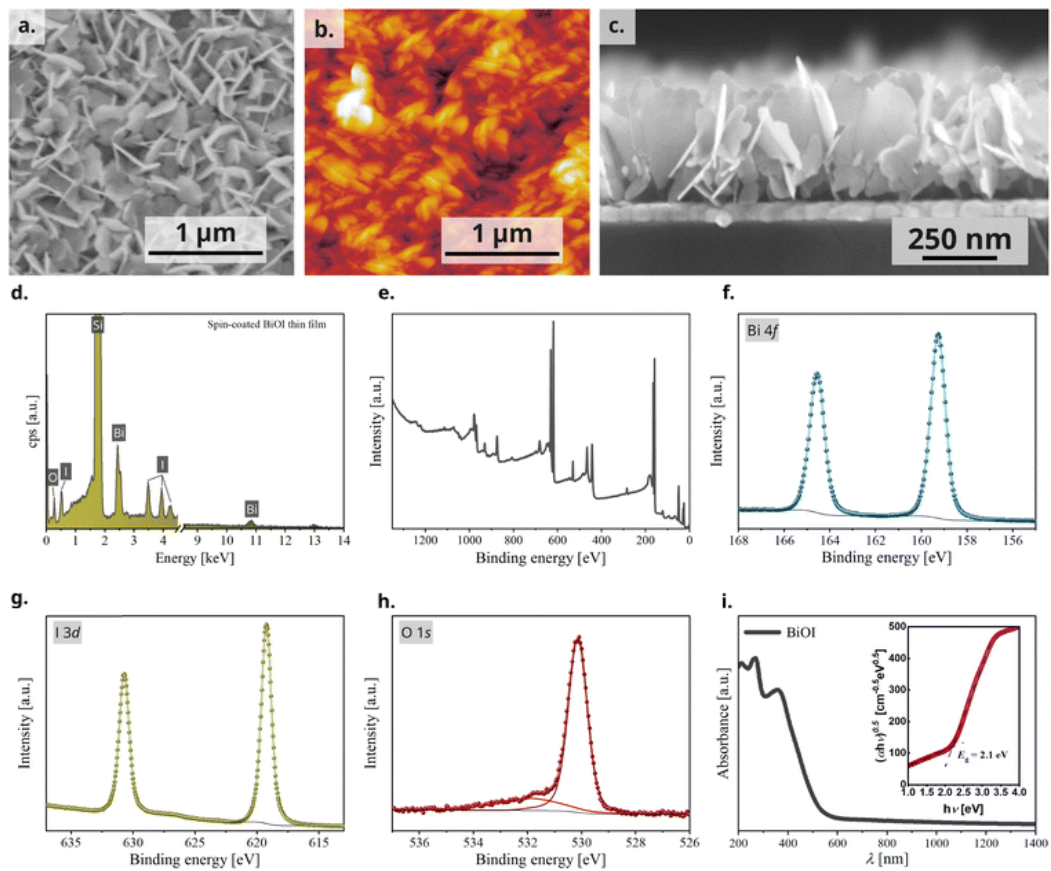

**Fig. 3** For BiOI thin film, (a) top-view SEM image, (b) AFM image, (c) cross-sectional SEM image, (d) representative EDX spectrum, (e) XPS overview spectrum, high resolution spectra of (f) bismuth, (g) iodine, and (h) oxygen and (i) UV-vis-NIR absorption spectrum (inset: Tauc plot).

To further probe the oxidation states and the chemical environment of the elements on the surface of the BiOI film, we performed X-ray photoemission spectroscopy (XPS) (survey spectrum, Fig. 3e). We observed one doublet peak for Bi 4f and I 3d, respectively (Fig. 3f and g). The binding energies of 159.3 eV for Bi $4f_{7/2}$ and 619.2 eV for I $3d_{5/2}$ can be assigned to bismuth and iodine in BiOI and are in agreement with literature.[37,41] In the case of O 1s, we observed two singlet peaks (Fig. 3h). The O 1s peak at 530.2 eV can be assigned to oxygen in BiOI, while the peak at 531.9 eV is associated with carbonyl and hydroxyl species (CO and $OH^-$) that could be present on the surface of the film due to fabrication in an aqueous environment and annealing in ambient air.

Lastly, we characterized the light absorption properties of spin-coated BiOI thin films and the UV-vis-NIR absorption spectrum shows a peak at 430 nm and an absorption cut-off at 525 nm. Tauc transformation of the absorption spectrum gives an indirect bandgap of 2.1 eV, which is also consistent with the orange-red appearance of the film (Fig. 3i).

**Characterization of OSC/BiOI bilayers**

Hybrid OSC/BiOI bilayers were fabricated as elaborated in the experimental section with BiOI spin-coated on top of shear-coated OSC layer. The OSCs chosen for investigation here are the conjugated polymers **P3HT**, **PCDTPT**, **DPPDTT**, **PDPP4T** and the small-molecule OSC **TIPS-pentacene** (Fig. 1) to include a wide range of possible applications with BiOI as a heterojunction, such as in solar cells, sensors and photoresponsive electronics.[42,43] Conditions for coating the OSC layer were optimized to obtain a closed and uniform thin film. Furthermore, solution shearing was used as it is generally a greener method than vapor techniques or other solution coating techniques and is also reported to result in pre-aligned films with better transport properties.[44,45] Since OSC parameters such as molecular *vs.* polymeric nature, chemical structure, surface energy of the film, *etc.*, have a strong influence on the heterojunction between the organic compound and the BiOI overlayer and consequently on the hybrid device, we first studied these bilayers with respect to the surface characteristics of the OSCs.

As seen in the SEM images, the platelet morphology of the BiOI particles in each of the hybrids remains the same as in the pristine BiOI film (Fig. 4). The alignment of particles and the morphology and density of the film in the case of **DPPDTT**, **PDPP4T** and **TIPS-pentacene** also remains indistinguishable under the SEM compared to the pristine BiOI film. In the case of **P3HT**, the particles appear to agglomerate to some extent to result in flower-like morphology of the film with large voids in between. For **PCDTPT**, there are areas on the film where particles are aligned horizontally (Lower magnification image, Fig. S2, ESI†), unlike the rest of the film where they are predominantly vertical. The SEM images also prove that this fabrication method for the BiOI thin film is suitable for obtaining its heterojunctions with the OSCs in a sequential fashion.

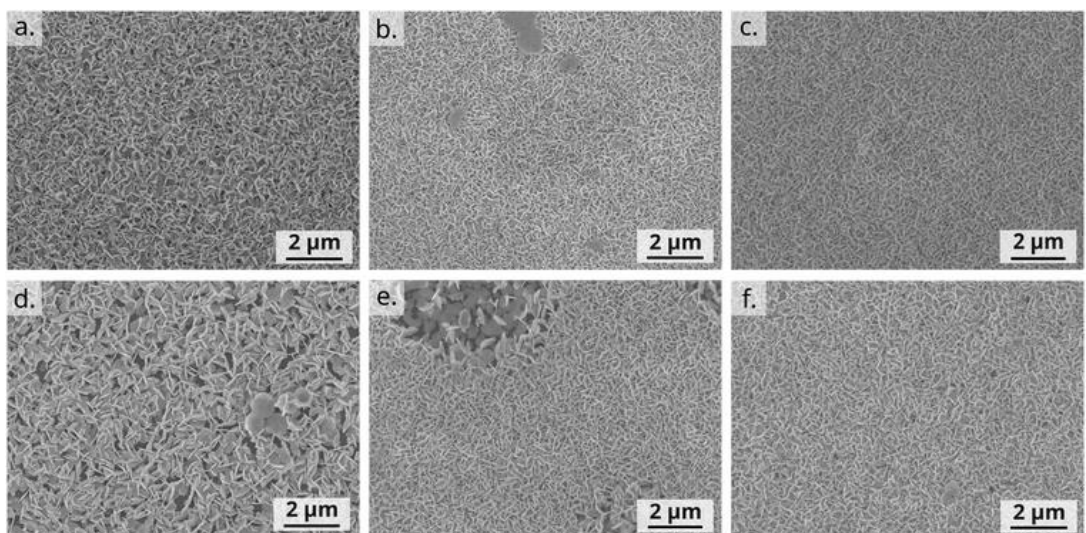

**Fig. 4** Top-view SEM images of (a) pristine BiOI and spin-coated BiOI hin films on shear-coated (b) **DPPDTT**, (c) **PDPP4T**, (d) **P3HT**, (e) **PCDTPT** and (f) **TIPS-pentacene** OSC films.

Furthermore, atomic force microscopy AFM was performed in non-contact tapping mode on pristine materials and heterojunctions to study the morphology and measure the RMS roughness of the films (Fig. S3 and S4, ESI†). Pristine OSC films show similar values of RMS roughness of about 1.25 nm, with the exception of **TIPS-pentacene** for which it is about 15.5 nm. This is due to the tendency of **TIPS-pentacene** to crystallize rapidly out of the coating solution into large grains as seen in the AFM image. The thickness of pristine materials and heterojunctions was measured using profilometry. On plotting the thickness of BiOI layers and RMS roughness of the corresponding OSCs, we see that there is no significant correlation of BiOI film thickness with the roughness of the underlying OSC film (Fig. 5a).

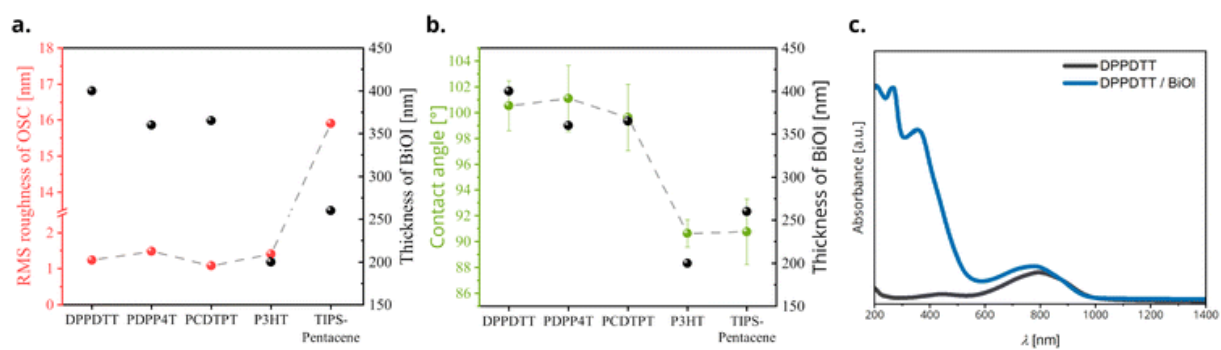

**Fig. 5** (a) RMS roughness of OSCs and thickness of the BiOI layer plotted for the various OSCs, (b) water contact-angle of OSCs and thickness of the BiOI layer for the corresponding OSCs and (c) UV-vis-NIR absorption spectrum of a representative **DPPDTT**/BiOI bilayer stack.

To analyze the influence of the OSC surface energy on the BiOI layer formation, we performed water contact-angle measurements on the pristine OSC films. Here, we clearly observe that although all the OSCs used are hydrophobic in nature, even small variations in the surface hydrophobicity have a pronounced effect on the thickness of BiOI films, with OSCs having a higher water contact-angle (hydrophobic) promoting growth of thicker films (Fig. 5b). Considering that the BiOI film was fabricated using spin-coating from a polar organic solvent (tetrahydrofuran, THF), this is to be expected. A lower contact angle of the precursor solution corresponds to a stronger interaction with the surface, which means that the solution spreads thin and spins out faster and results in a thinner film. On the other hand, a higher contact ensures the precursor droplets retain a more spherical shape due to surface tension, and this combined with the high volatility of THF, results in a thicker film.

We studied the light absorption behavior of the bilayer heterojunctions using UV-vis-NIR spectroscopy. Absorption spectra of all the OSC/BiOI bilayers show superposition of the spectra of their components, demonstrating that both components are chemically intact. Since the ranges of strong absorption are largely complementary, most of the heterojunctions tested here effectively cover the UV-vis-NIR region with the absorption cutoff depending on the choice of the OSC. While BiOI shows an absorption from UV to green light (cut-off at 525 nm), most of the OSCs used here extend the absorption to the NIR region (Fig. S5, ESI†). The absorption spectrum of one representative stack **DPPDTT**/BiOI is given in Fig. 5c. This paves way for a prospective application of these heterojunction films in photoresponsive devices. Although it is beyond the scope of the current paper, this is an interesting aspect of these heterojunctions yet to be explored.

**OSC/BiOI heterojunction FETs**

After evaluating the structure, composition, and morphology of the BiOI thin film in depth and characterizing the OSC/BiOI heterojunctions, we integrated them into field-effect transistor devices to study their electrical characteristics. The configuration chosen for this was the bottom-gate bottom-contact (BGBC) architecture comprising of $SiO_2$ dielectric, Cr/Au contacts followed by the OSC and BiOI layers (schematic representation, Fig. 6a). Bottom contact architecture was preferred here to ensure an efficient injection/extraction into/from the OSC as the charge transport material. The cross-sectional SEM image of one representative heterojunction device, **DPPDTT**/BiOI (Fig. 6b), shows the device configuration with all the layers. It is clearly visible that this architecture is most promising as the OSC layer ensures the adhesion and contact of the highly textured BiOI film to complete the heterojunction device.

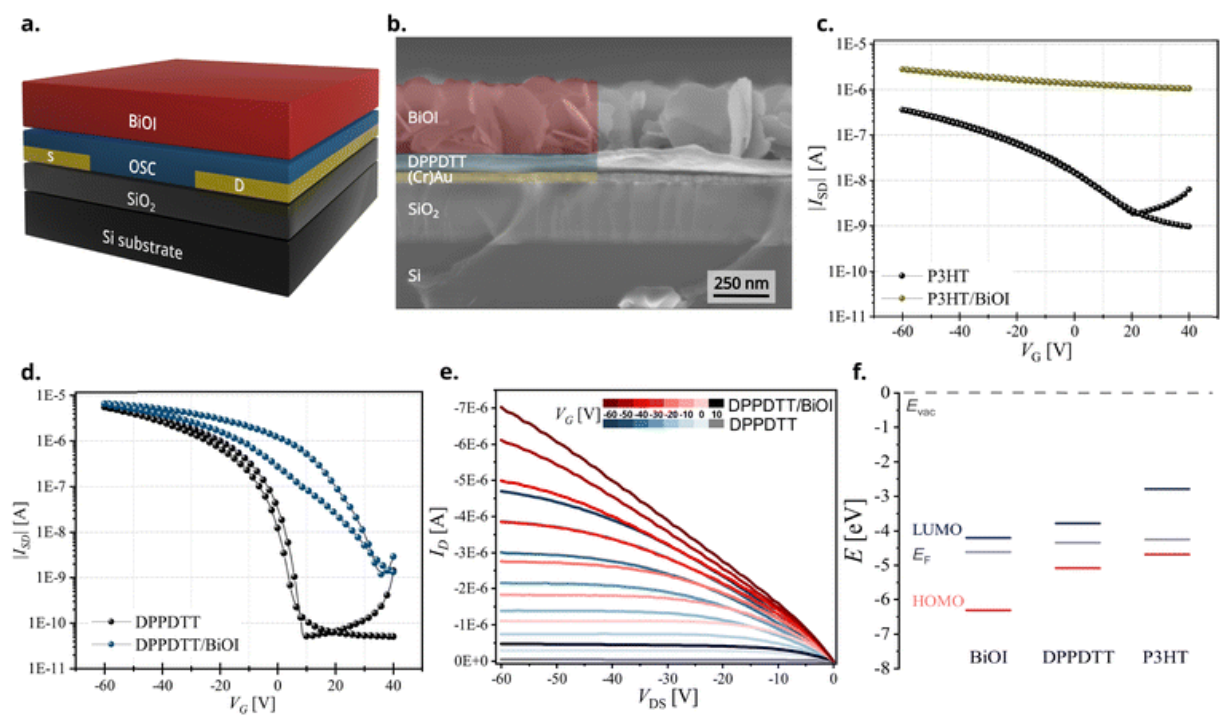

**Fig. 6** (a) Schematic representation of the device architecture, (b) CS-SEM of a representative **DPPDTT**/BiOI heterojunction device, transfer characteristics of pristine OSCs and heterojunction devices ($V_{DS}$ = −60 V) for (c) **P3HT** and (d) **DPPDTT**, (e) Output characteristics of pristine and heterojunction devices with **DPPDTT** in a range of gate voltages from 10 to −60 V and (f) energy level diagram of BiOI and OSCs **DPPDTT** and **P3HT** thin films determined using UPS and UV-vis-NIR spectroscopy.

By analyzing the transfer characteristics of the heterojunction devices and the control (pristine OSC) devices, we aim to identify the most suitable OSC/BiOI heterojunction for applications employing the FET configuration and understand the key design parameters influencing the performance of these heterojunction devices. Distinct device characteristics are immediately apparent in each hybrid heterojunction. These distinctions likely arise from both, physical modification of the OSC layer and changes in their electronic properties at the heterojunction interface. Transfer characteristics of heterojunction FETs with all conjugated polymers display strong oxidative doping of the polymers by the BiOI overlayer. Despite confirmation of heterojunction formation of BiOI with **TIPS-pentacene** from UV-vis spectroscopy and SEM, the heterojunction devices showed no conductivity (Fig. S6a, ESI†). One possible reason could be the partial solubility of **TIPS-pentacene** in the processing solvent of BiOI, THF, which can trigger unfavorable changes in the macroscopic stacking in the film upon introduction of the rigid BiOI overlayer.[46]

Heterojunction devices employing **DPPDTT** and **PDPP4T** exhibit FET behavior with moderately increased on-currents compared to the organic-only devices, while devices with **P3HT** and **PCDTPT** become conducting and lack sufficient gate control. Hybrid FETs with **P3HT**, a donor-type conjugated polymer (Fig. 6c and Fig. S7a, ESI†), or **PCDTPT**, a donor–acceptor conjugated polymer (Fig. S6b and S7b, ESI†), show drastic shifts in device performance in the heterojunction devices leading to an always-ON behavior with very low ON/OFF ratio. Even though these heterojunctions are not suitable to be employed in transistors, they are interesting to follow up for devices employing BiOI that require charge (hole) extraction layers, such as solar cells.[47]

**DPPDTT** and **PDPP4T**, both donor–acceptor conjugated polymers, are most interesting for heterojunction devices as they are well known for their high mobility and ambient stability.[48] It is found that BiOI heterojunction devices with these polymers indeed show a transistor behavior with good ON/OFF ratio (Fig. 6d and Fig. S6c, ESI†). They show a shift in the threshold voltage (Fig. S7c and S7d, ESI†) and an improved ON-current compared with pristine OSC devices. The output characteristics swept for a range of gate voltages from 10 V to −60 V for the **DPPDTT**/BiOI heterojunction and control (**DPPDTT**) FETs further show this increase in the ON-current (Fig. 6e). The electrical performance parameters related to FETs, mainly the field effect mobility ($\mu_{av}$), threshold voltage ($V_{TH}$) and the ON/OFF ratio for the heterojunction devices are summarized in Table 1. The devices with **PDPP4T** exhibit a lower ON/OFF ratio compared to those with **DPPDTT**, and they both show a shift in the $V_{TH}$ relative to the pristine OSC devices.

**Table 1** Measured transistor parameters for different bilayer heterojunction devices (average of 5 devices)

| Sample | $V_{TH}$ [V] | $\mu_{av}$ [$10^{-4}$ cm$^2$ Vs$^{-1}$] | ON/OFF ratio |
|---|---|---|---|
| **P3HT** | 30 ± 2 | 1.3 ± 0.2 | $2 \times 10^2$ |
| **P3HT**/BiOI | — | 7.1 ± 0.6 | $1 \times 10^1$ |
| **PCDTPT** | 8 ± 5 | 2.2 ± 0.6 | $4 \times 10^4$ |
| **PCDTPT**/BiOI | — | 2.0 ± 0.1 | $1 \times 10^1$ |
| **DPPDTT** | 2 ± 2 | 75.5 ± 0.3 | $1 \times 10^5$ |
| **DPPDTT**/BiOI | 28 ± 4 | 60.1 ± 0.4 | $5 \times 10^3$ |
| **PDPP4T** | 1.5 ± 0.5 | 100 ± 4 | $1 \times 10^4$ |
| **PDPP4T**/BiOI | 6.6 ± 1.1 | 110 ± 9 | $2 \times 10^2$ |

One known reason for a shift in the threshold voltage and increase in the OFF-current is charge transfer (doping), here occurring between the OSC and the inorganic BiOI layer. Despite the similar positions of the highest occupied molecular orbital (HOMO) level of the OSCs as calculated from UPS and UV-vis spectroscopy (Table S1 and Fig. S8, ESI†), we see significantly different FET device characteristics in their hybrids. We also note that the fabricated BiOI thin film is n-doped, which was evidenced in a few previous reports.[40,41] This unusual p-doping of the OSCs by BiOI and the differences in device behavior among the different conjugated OSCs investigated here can be best discussed by taking into account the pronounced thiophilicity of bismuth and its implications when interfaced with polythiophenes. The high affinity of bismuth for sulfur, like other heavy metals (lead, gold, *etc.*) gives rise to the possibility of partial oxidation of the polymer films by the bismuth(III) cation in BiOI *via* charge transfer from the 3*p* lone pair of the S atom to the (formally) empty 6p orbitals of the bismuth(III) atom. This has been shown before in the case of **P3HT** as contact doping at a **P3HT**–Au interface with an oxidation of up to 20% of the thiophene units of the first monolayer at the interface and also charge-transfer complexes of **P3HT** with molecular oxygen.[49,50]

The underlying chemical principle can be extended to the systems studied here. These interactions result in the formation of charge-transfer complexes causing an oxidative p-doping of the polymers, which is reflected in the device behavior of all conjugated polymer heterojunctions investigated here. The extent of this doping depends predominantly on the chemical nature and energetic position of frontier levels of the organic and inorganic layers and the area of interface at the heterojunction. The electronic charge transfer from the HOMO level of the OSC to the bismuth(III) states, which form the conduction band minimum (CBM) in BiOI, results in (additional) holes in the OSC valence band. These heteropolar interfacial interactions are especially efficient in the case of BiOI due to its very deep lying conduction band and relatively exposed bismuth atoms at the crystal surfaces. OSCs with a higher HOMO level, *i.e.*, **P3HT** and **PCDTPT**, show a higher propensity towards oxidation by BiOI compared to OSCs with deeper lying HOMO levels, *i.e.*, **DPPDTT** and **PDPP4T** (Fig. 6f). In **DPPDTT** and **PDPP4T**, this oxidative doping of the film also limits the ability of the active device layer to fully deplete at positive gate bias voltages, which is noted as an increased OFF-current in the heterojunction devices compared to pristine OSC devices. On the other hand, we find that the average mobility remains comparable to the pristine counterparts in these cases (Table 1).

The surface termination of a free-grown BiOI platelet in the stacking direction *i.e.*, the main facet of the crystal (001), must ensure charge balance and would ideally be the iodide layer. However, due to the processing in water, hydroxide ions can partially replace the iodide ions on the surface, which is noted here in the XPS analysis. Heteropolar bismuth–sulfur interactions can thus broadly occur in two ways: either on the main faces of the BiOI platelets, involving chemical substitution of iodide or hydroxide by coordinating sulfur atoms of the OSC, or on the BiOI edge facets by direct coordination of the sulfur atoms to the bismuth(III) cations (Fig. 7). However, the density of bismuth atoms on (001) is about 2.3 times higher than on (100). Given the same contact area with OSC, the charge transfer should be higher at the interface with the (001) face. Moreover, we observe excessive doping for the sulfur-rich **P3HT** and **PCDTPT**. The comparatively medium concentration of accessible sulfur donors in **DPPDTT** and **PDPP4T**, however, proves to be suitable. All these aspects highlight another key parameter that determines the oxidative p-doping of the OSCs by the BiOI layer, namely the density and availability of chemically interacting sulfur and bismuth atoms at the interface. The case of BiOI heterojunction with **PCDTPT**, for which we see systematically large areas of the film where the particles have grown horizontally (Fig. S2, ESI†), also hints towards the importance of interfacial area for device characteristics.

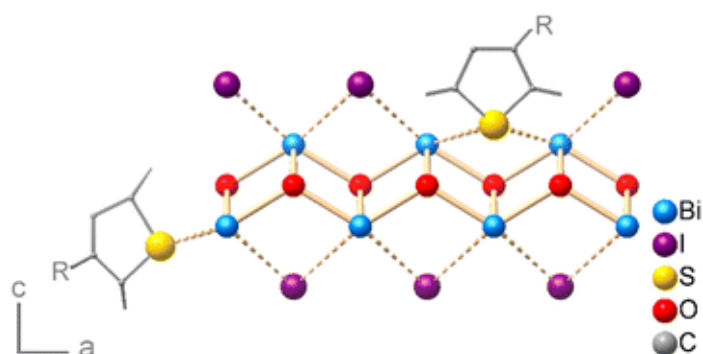

**Fig. 7** Schematic visualization of possible bonding situations at the OSC/BiOI interface (side or main face contact) for a generalized (poly)thiophene.

To demonstrate the importance of the area of interface at the heterojunction, we perform a control experiment by shear-coating a second layer of **DPPDTT** onto the original **DPPDTT**/BiOI heterojunction. Considering the flake-like texture of BiOI, we expected the second **DPPDTT** layer to percolate in the voids of the BiOI layer and cling around the BiOI platelets thus having a much higher interfacial area with them and consequently show an always-ON device behavior similar to **P3HT** and **PCDTPT** heterojunction FETs. The increased interface between BiOI and the second **DPPDTT** film is evident in the top-view and cross-sectional SEM images (Fig. 8a and b), and we confirm the expected excess oxidation of the **DPPDTT** film by measuring the transfer characteristics of the device so formed (Fig. 8c). This is further substantiated by measurement of two terminal devices and their current–voltage ($I$–$V$) characteristics (Fig. 8d). This proves that even in the case of OSCs like **DPPDTT** and **PDPP4T**, which are less prone to undergo oxidative p-doping due to their deeper HOMO levels, it is of utmost importance to control the area of the interface between the two layers at the heterojunction in order to control the extent of doping of the OSC layer. Thus, the vertical growth of BiOI particles in the film is in fact crucial to the electrical characteristics of the heterojunction devices with **DPPDTT** and **PDPP4T** as it caps the extent of doping of the OSC layer, while at the same time allowing the integration of the functionalities of BiOI into a functional device. A second experiment was performed to further establish this. The morphology of the BiOI thin film was tuned by changing the concentration of the precursor BiI$_3$ solution. On using a lower concentration (0.05 mol L$^{-1}$) of the BiI$_3$ solution during spin-coating, predominantly horizontally grown particles could be obtained in the BiOI thin film (sample **A**) in contrast to the one grown from a 0.2 mol L$^{-1}$ BiI$_3$ solution (sample **B**) (SEM image, Fig. S9a and b, ESI†). This resulted in a higher area of interface with the OSC in the bilayer heterojunction and consequently, higher oxidative doping of the **DPPDTT** polymer in the hybrid FET devices (transfer curves, Fig. S9c, ESI†). We note a one-order increase in the OFF current in sample **A** compared to sample **B**, with sample **A** showing a $V_{TH}$ of 40 V. The higher number of free charges in the hybrid FET fabricated from 0.05 mol L$^{-1}$ solution of BiI$_3$ results from an increased oxidative doping due to a higher interfacial area between the OSC and the BiOI layers as compared to the 0.2 mol L$^{-1}$ counterpart, and as such provides a direct evidence of the argument presented here.

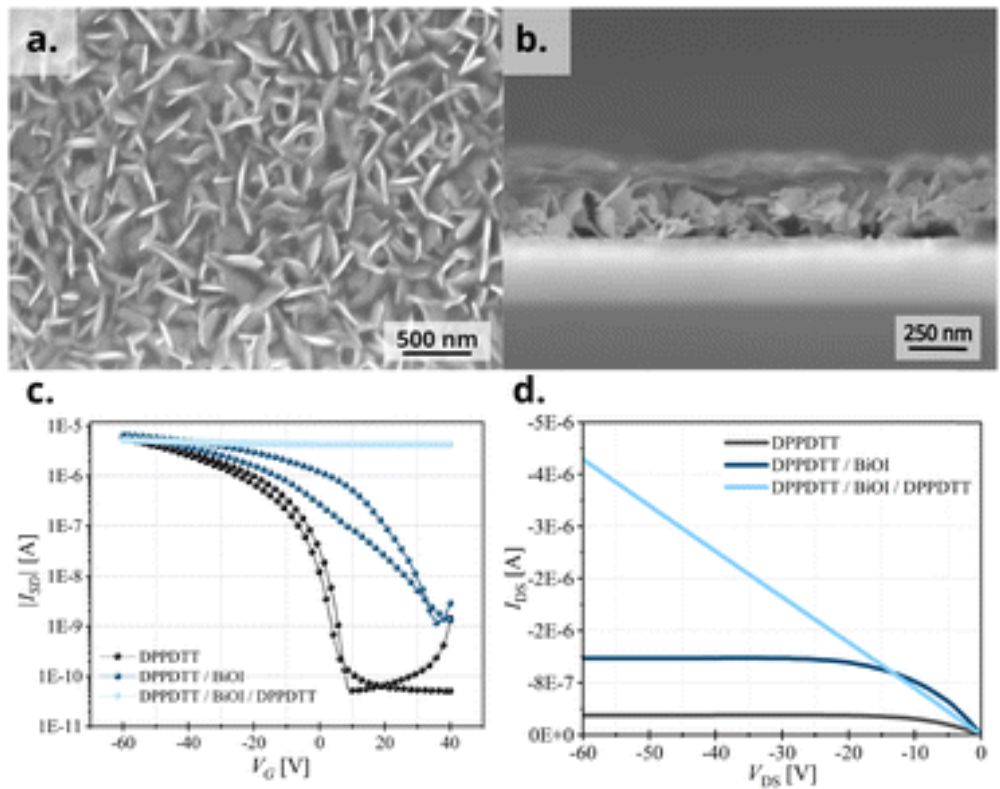

**Fig. 8** For three-layer **DPPDTT**/BiOI/**DPPDTT** device, (a) top-view SEM and (b) cross-sectional SEM confirming a higher area of interface, (c) the transfer characteristics and (d) $I$–$V$ curves as compared with the bilayer heterojunction and pristine OSC device.

Since, charge transfer processes at the heterojunction are related both to the chemical and physical properties of the OSC surface, we evaluated the influence of these properties on the transistor performance using two important features of the OSC layer, the surface roughness and the water-contact angle. When we revisit the plot of RMS roughness (Fig. 5a) of the OSC films against the investigated OSCs on the *x*-axis, we see that even though there is no rigorous correlation between OSC roughness and device performance, OSCs for better performing heterojunctions, *i.e.*, ones that have the desired FET characteristics, have smoother surfaces. On the other hand, the water contact-angle on OSC surfaces (Fig. 5b) shows a strong correlation with the heterojunction FET device performance, with OSCs having a higher contact-angle *i.e.*, **DPPDTT** and **PDPP4T**, outperforming others. This is to be expected since a surface of high hydrophobicity withstands the water-based processing of the BiOI layer much better, which translates to an enhanced device performance. Lastly, we measured the stability of the fabricated **DPPDTT**/BiOI heterojunction FETs. The devices retain their electrical characteristics for at least six months under ambient conditions and exhibit high ON/OFF cycling stability, measured for 100 cycles after six months (transfer curves, Fig. S10, ESI†). They show a slight increase in the OFF current (around +10 nA) and shift of +10 V in the $V_{TH}$ with storage. This can be attributed to doping of the OSCs by atmospheric oxygen over time, which has been reported before.[44,50,51] The high ambient and cycling stability of the hybrid BiOI FETs as compared to pure halide materials, for example bismuth- or lead-based halide perovskites stems also from the high ambient resilience and limited halide migration in BiOI.[51–53]

We thus show that with the correct choice of material combination, here the conjugated OSCs **DPPDTT** and **PDPP4T**, and an in-depth understanding of the influencing parameters, we can successfully integrate solution-processed BiOI into functional lateral transport devices.

**Conclusion**

We demonstrate, for the first time, organic–inorganic hybrid FET devices employing the stable, environmentally benign and non-toxic material BiOI. As-obtained solution-processed thin films were utilized by combining them with high performing p-type organic semiconductors. Fabrication of the heterojunction FET devices was facilitated without any intensive processing techniques or employing any chemical or physical modification agents. We identify heterojunctions of BiOI with **DPPDTT** and **PDPP4T** to be most suitable for applications based on the FET architecture. Importantly, we provide insights into the complex nature of such hybrid systems by discussing a broad range of interactions possible at the heterojunction that led to the unique device behaviors for each of the investigated OSCs. We highlight the intricate interplay of the physical and chemical characteristics of the individual components, most importantly the position of frontier energy levels and the area of interface at the heterojunction of these hybrids. These identified parameters can also be used for tuning the performance of devices based on BiOI and establish a foundational design strategy for selection of the active components in heterojunctions employing it. In effect, we establish that such organic–inorganic hybrid systems provide the ideal controls to harvest respective material properties synergistically. In this case, BiOI with its favorable optical and chemical properties brings in new functionalities to the FET devices, and the conjugated OSC **DPPDTT** with its superior charge transport properties compared to BiOI and smooth interfaces enables a functional device. The heterojunction hybrid systems investigated here now open avenues for investigation of these combinations for specific applications including but not limited to photodetectors, phototransistors, sensors, and solar cells.

### Author contributions

VL and PD contributed equally to this work. MR did the conceptualization, supervision, and provided the resources and funding for the work. VL was involved in development of methodology, data curation and analysis, visualization, and writing the original draft. SCBM provided supervision, resources, and funding for the work. PD was involved in development of methodology, data curation and analysis, and writing original draft. MH performed GIWAXS data acquisition and analysis. YJH performed the XPS and UPS data acquisition and analysis under the supervision of YV. All authors contributed to the editing and reviewing of the manuscript.

### Acknowledgements

The GIWAXS experiments were performed at BL11-NCD-SWEET beamline at ALBA Synchrotron with the collaboration of ALBA staff. We thank Marc Malfois for help with setting up the experiment. This work received financial support from the German Research Foundation (DFG) through the Collaborative Research Center "Chemistry of Synthetic Two-Dimensional Materials" (SFB 1415, project-ID 417590517), the Würzburg-Dresden Cluster of Excellence on "Complexity and Topology in Quantum Matter – ct.qmat" (EXC 2147, project-ID 390858490) and the European Research Council (ERC) under the European Union's Horizon 2020 research and innovation programme (ERC Grant Agreement no 714067, ENERGYMAPS). Open Access funding provided by the Max Planck Society.